\documentstyle[fullpage,graphics,twocolumn]{article}
\newcommand{\PSbox}[3]{\mbox{\rule{0in}{#3}\includegraphics{#1}\hspace{#2}}}

\title{
A Multiscale Approach to Determination of Thermal Properties
and Changes in Free Energy: Application to Reconstruction of
Dislocations in Silicon
}

\author{T.D. Engeness and T.A. Arias\\
Department of Physics\\
Massachusetts Institute of Technology
Cambridge, MA 02139}

\begin{document}

\maketitle

\begin{abstract}
We introduce an approach to exploit the existence of
multiple levels of description of a physical system to radically
accelerate the determination of thermodynamic quantities.  We first
give a proof of principle of the method using two empirical
interatomic potential functions.  We then apply the technique to feed
information from an interatomic potential into otherwise inaccessible
quantum mechanical tight-binding calculations of the reconstruction of
partial dislocations in silicon at finite temperature.  With this
approach, comprehensive {\em ab initio} studies at finite temperature
will now be possible.\\
{\bf 65.50.+m,61.72.Lk,02.70.Lq}
\end{abstract}

Our understanding of the detailed processes underlying the behavior of
condensed matter systems has progressed tremendously over the last
fifteen years.  Electronic structure techniques give insight into the
electronic origins of phenomena, but most readily for fixed lattices
at zero temperature.  Modern simple, but realistic, interatomic
potentials reveal the lattice processes underlying complex phenomena
at finite temperature.  One of the greatest challenges in condensed
matter theory is to understand the electronic origins of solid state
phenomena at finite temperature.  Comparatively little progress has
been made to date in this area because of the tremendous complexity of
combining the large number of degrees of freedom needed to describe
the electrons with the large number of ionic positions needed to
describe the occupied regions of the lattice phase space at finite
temperature.  In this letter we show that there is an exact
separation of the finite-temperature electronic-structure problem into
two far simpler parts, a finite-temperature lattice part, which may be
studied using approximate coarse-grained lattice models, and the
electronic structure part, which may be studied using highly accurate,
fine-grained calculations.

Empirical interatomic potentials do not treat the electronic
degrees of freedom explicitly, but have proved successful in modeling the
general behavior of materials ranging from insulators \cite{TerC}
through semiconductors \cite{SW,Ter} and metals \cite{Daw12}.
These models capture basic interatomic behavior and their simplicity
makes them well suited for the evaluation of thermal averages.
However, because empirical models coarse-grain over the
electronic degrees of freedom, they generally do not provide
sufficiently accurate information for quantitative predictive studies.

Tight-binding models represent a next
step in detail of description and reliability.
These models include the electrons explicitly, but restrict their wave
functions to linear combinations of atomic
orbitals \cite{Harrison,Sawada,Kohyama}.  These potentials are
therefore applicable over a wider range of phase space than interatomic
potentials and provide certain electronic information.  Tight-binding
calculations are far more demanding than their empirical interatomic
counterparts.
Direct Monte Carlo or molecular dynamics calculations of systems
at finite temperature are sufficiently demanding that the development of
specialized approximate techniques is an active area of
research \cite{Krajci}.

{\em Ab initio} calculations attempt to describe all relevant
electronic degrees of freedom.  They give a good description of the physics of
condensed matter systems over a wide range of phase space \cite{RMP}.
At present, density functional based calculations represent the
greatest level of detail at which
extended crystalline defects may be studied.
Because the calculation of thermal averages with Monte
Carlo or molecular dynamics methods requires the evaluation of many
configurations to sample phase space fully, {\em ab 
initio} calculations of thermal properties such as free energy
differences \cite{Milman} are tremendously demanding and infrequently
attempted.

Finite temperature studies require a strategy for
obtaining a proper ensemble.
In general, the populated regions of phase
space consist of a very narrow surface.  To explore this surface with
uniform sampling is impossible for a complex system because the
surface occupies such a miniscule volume of phase space.  As an
alternative, one may start from a point on the occupied surface and
then make small steps to explore it.  One then always
explores relatively relevant points in phase space, but covering the
entire surface then requires an enormous number of steps.
This can be
manifested as correlation between large numbers of consecutive samples
or as becoming trapped in local energy minima.
This
small-step approach is the essence of the two standard methods for
evaluation of thermal averages: molecular dynamics \cite{MD} and Monte Carlo
calculations based on the Metropolis algorithm \cite{Me}.

As an alternative, we propose to separate the problem into two parts.
First, we fully explore the relevant regions of phase space using a
coarse-grained Hamiltonian, such as an interatomic potential for which
extensive calculations based upon one of the traditional small-step
approaches are tractable.  Once the relevant regions
are identified, we evaluate physical observables
within the target fine-grained Hamiltonian, which may include
electronic information, and combine the results to
obtain proper averages over the fine-grained ensemble.
Below we demonstrate that modern interatomic potentials are sufficiently
close to their {\em ab initio} counterparts that thermal averages may
be computed to within the accuracy of density functional theory by
performing {\em ab initio} calculations on a very limited number of
samples drawn from the interatomic ensemble.  Our approach takes optimal
advantage of the radical separation in computational time scales which
interatomic potentials and quantum mechanical calculations exhibit.
Rather than employing {\em ad hoc} non-physical Hamiltonians in a
spirit similar to umbrella sampling \cite{umbrella}, we propose the
use of physical, albeit coarse-grained, Hamiltonians.

\def\cH{{\mathcal{H}}}
\def\cO{{\mathcal{O}}}

{\em Corrected Ensemble Averages} --- The first phase of our approach
is to run a large-scale exploration of phase space with a simplified
model, using an appropriate, traditional thermodynamic approach.
Drawing samples from the resulting configurations allows the more
demanding model to be applied to truly uncorrelated, properly
distributed points in phase space.  So long as the two Hamiltonians
are sufficiently correlated, we are assured that the samples selected
are physically relevant to ensemble averages over the fine-grained
Hamiltonian.

To correct for the differences in the thermal distributions of the two
Hamiltonians, each sample $i$ selected from the initial simulation
must be weighted by its relative probability in the two
ensembles, the {\em corrective Boltzmann factor}
\begin{equation}
B^{f \leftarrow c}_i = \exp \left [-\beta ( \cH^f_i - \cH^c_i ) \right ].
\end{equation}
Here, $\beta$ is the inverse temperature, and $\cH^c_i$ and $\cH^f_i$
are the energy of configuration $i$ within the coarse-grained and
fine-grained models, respectively.
Once given the corrective Boltzmann factors, the average of any observable
${\cO}$ in the ensemble of $\cH^f$ is
\begin{equation} \label{eqn:aveO}
<{\cO}>_f =\frac{\sum_i
B^{f \leftarrow c}_i {\cO}_i}{\sum_i B^{f \leftarrow c}_i}.
\label{eq:newav}
\end{equation}

{\em Free energies} --- One common approach to the calculation of
free energies is thermodynamic integration \cite{FI}, which gives the
free energy difference between two macrostates
characterized by the parameter values $\lambda=0$ and $\lambda=1$ as
$
\Delta F = \int^1_0 \left < \frac{\partial E}{\partial \lambda}
\right >_{\lambda} d\lambda,
$
where the average is computed over
the Boltzmann distribution at fixed $\lambda$.
In practice, the integral is computed numerically by
sampling a finite number of values of $\lambda$.  When
$\lambda$ is a generalized coordinate, $\Delta F$
is the integral of the thermal average of the corresponding
generalized force.

{\em Proof of Principle} --- To demonstrate the soundness of this
approach we use two potentials which are sufficiently simple to allow
direct comparisons using brute-force techniques.  We have
chosen two models which differ from each other at least as much as
typical interatomic potentials differ from density functional calculations.
This allows us to address the issue of whether or not empirical
potentials yield distributions sufficiently close to {\em ab initio}
distributions that (\ref{eqn:aveO}) may be evaluated reliably using a
tractable number of samples.

To play the role of the coarse-grained Hamiltonian $\cH^c$, we have
chosen an early version of an interatomic potential for silicon which
is still under development.  The Stillinger-Weber model \cite{SW} plays
the role of the fine-grained Hamiltonian $\cH^f$.  The standard
deviation in energy differences $\cH_f-\cH_c$ over thermally
distributed samples provides a useful measure of the inter-Hamiltonian
discrepancy.  We found that, for the system we study below, this
measure is the same (to within ten percent) for our two interatomic
potentials as it is for the Stillinger-Weber potential and
density-functional calculations.  Later we shall apply this measure to
determine the relative suitability of different interatomic potentials
in our procedure.

\begin{figure}
\begin{center}
\scalebox{0.35}{\includegraphics{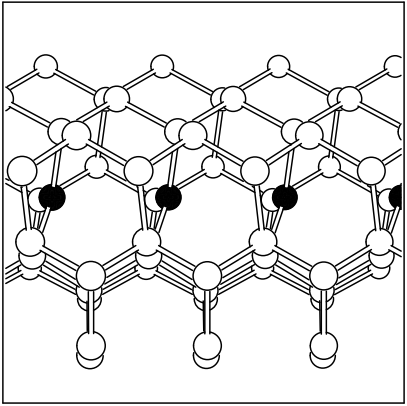}}
\end{center}
\caption{Unreconstructed ($\lambda=1$) configuration of the 30$^o$ partial
dislocation (glide set) in silicon.  Atoms in black are core atoms
which bond together in pairs upon reconstruction.}
\label{fig:disloc}
\end{figure}

The physical system to which we apply our method
is the 30$^o$ partial dislocation (glide set) in silicon.
Figure~\ref{fig:disloc} shows the core of this dislocation in the
unreconstructed state.  In nature, the core atoms indicated in black
in the figure move together in a period-doubling reconstruction to
form dimers which saturate all dangling bonds and thereby minimize
the energy of the dislocation.  To illustrate our approach, we
calculate the free energy of this reconstruction at $T=930$K, where
some indirect experimental information is available
 \cite{George_Champier,Alexander}.  We use a periodic supercell of 96
atoms containing a dislocation-antidislocation core at a separation of
14\AA.

\begin{figure}
\begin{center}
\PSbox{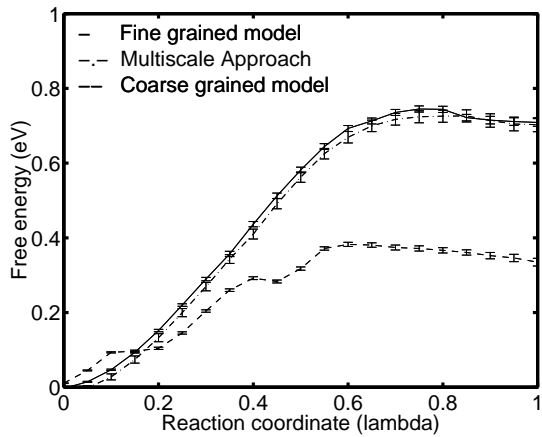 hoffset=-15  voffset=-80 vscale=40 hscale=40}{3.3in}{2.25in}
\end{center}
\caption{Free energy of the silicon 30$^\circ$ partial dislocation
(glide set) as driven from its reconstructed ($\lambda=0$) to its
unreconstructed ($\lambda=1$) state.}
\label{fig:comfi}
\end{figure}

Figure~\ref{fig:comfi} shows the cumulative changes in free energy for
$\cH^c$ and $\cH^f$ (dashed and solid lines, respectively) as a
function of $\lambda$ as we drive the dislocation core from its
reconstructed ($\lambda=0$) to its unreconstructed ($\lambda=1$)
state.  The figure also shows the statistical uncertainties remaining
after the evaluation of twenty-five and fifteen {\em million} samples for
$\cH^c$ and $\cH^f$, respectively.  As the figure illustrates, this
number of samples is necessary to produce a reliable estimate of the
final free energy difference within the Metropolis algorithm.  Note
that the final free energy change under $\cH^c$ is much lower than for
$\cH^f$ and that the energy profile for $\cH^c$ exhibits extraneous
local minima and maxima.  The challenge to our multiscale method
is to correct for the far lower free energy and spurious
local minima and maxima of $\cH^c$ while using
the {\em same} samples which led to the distorted curve
for $\cH^c$.

The dash-dotted line Figure~\ref{fig:comfi} displays the results we
obtain with the multiscale approach.  To produce these results, we
first made five separate runs each drawing only one {\em thousand}
independent samples at large intervals from the direct simulation
under $\cH^c$.  We then evaluate the generalized force using $\cH^f$
and compute the average force at each value of $\lambda$ for each run
according to (\ref{eqn:aveO}).  Finally, we find the mean and standard
deviation among the forces of the five runs at each $\lambda$ and
integrate the results numerically to give the free energy curve and
uncertainties that appear in the figure.  The fact that the newer
curve has quite similar statistical uncertainties to the previous
curves while computed from three orders of magnitude fewer samples
underscores the fact that the vast majority of samples in the
brute-force Metropolis approach only serve to generate the Boltzmann
distribution but do not contribute significantly independent
statistical information to the averages.  If a sampling scheme better
than the Metropolis algorithm were employed, the comparison would be
less favorable for our approach.  However, we note that our approach
is direct, efficient and as general as the availability of suitable
atomistic potentials.  Figure~\ref{fig:comfi} also shows that the
averaging procedure (\ref{eqn:aveO}) succeeds in eradicating the
spurious maxima and minima of the coarse-grained Hamiltonian.  The
final free energy of reconstruction determined using the multiscale
approach is $0.711\pm0.019$~eV, in good agreement with the result of
the brute-force method, $0.712\pm0.010$~eV.  We thus see that the low
free energy difference under $\cH^c$ is indeed properly compensated,
even when working with a radically reduced number of samples.

To obtain a statistical uncertainty in the free energy difference of
$0.043$~eV, which is within the accuracy of the best density
functionals \cite{Becke}, the preceding calculation could be done with
a single run, instead of five.  This would require the evaluation of a
total of only {\em one thousand} samples, quite feasible for {\em ab
initio} work.

While it is true that the corrective factors fluctuate exponentially
with the {\em total} energy discrepancy $\cH_f-\cH_c$, whose standard
deviation scales as the square root of the number of atoms in the
system, the corrected average (\ref{eqn:aveO}) does not fluctuate
among runs nearly as widely.  This is because (\ref{eqn:aveO}) is a
weighted average of the observable $\cO$, and, therefore, the
fluctuations of $\cO$ place an absolute upper bound on the
fluctuations of (\ref{eqn:aveO}).  The system size dependence of the
fluctuations of (\ref{eqn:aveO}) thus ultimately becomes the system
size dependence of the fluctuations of $\cO$.  If $\cO$ is a local or
intensive parameter, as we have here, these fluctuations do not
increase with system size.

For runs of fixed length, increases in fluctuations in the corrective
factors result in fewer values of $\cO$ contributing to the final
average.  It thus becomes more difficult to detect the correlation
between the observable and the inter-Hamiltonian discrepancy.
However, {\em local} observables, such as our generalized force, are
largely uncorrelated with the {\em global} energy discrepancy.  When
these quantities are completely uncorrelated, (\ref{eqn:aveO}) reduces
to $<{\cO}>_f=(1/N) \sum_i {\cO}_i$.  This direct average yields a
free energy of reconstruction of $0.682 \pm 0.008$~eV, in excellent
agreement with the exact result.  The correlation between the
generalized force and the energy discrepancy thus corresponds to the
remaining $0.02$ to $0.04$~eV in the free energy, which the corrective
Boltzmann factors capture rather well within our present run length
and system size.  Numerical experiments on model random variables
reveal that, for a fixed run length, the effect of correlation between
a local observable and the energy discrepancy degrades slowly with
system size, approximately as the square root of the number of atoms.
As the present scale of one hundred atoms approaches the maximum size
feasible for current {\em ab initio} calculations with extensive
exploration of phase space, we do not expect this degradation to
present a significant problem for some time.  Much can be done on the
scale of one hundred atoms with our approach in its present
formulation.

{\em Multiscale model of dislocation cores at finite temperature} ---
As a first truly multiscale application of our approach, we shall now
demonstrate the use of interatomic potential functions, which do not
deal explicitly with electronic degrees of freedom, to generate
samples for quantum-based tight-binding calculations, which do.  The
traditional Monte Carlo approach would demand the evaluation of
millions of tight-binding configurations, requiring years of computing
time.

For the present calculations, we use the tight-binding Hamiltonian of
Sawada \cite{Sawada} with the
modifications proposed by Kohyama \cite{Kohyama}.  This model
provides a relatively good description of the bulk, dimer and surface
energetics of silicon.  A limited number (about fifty) of preliminary
runs on our target system shows that the standard deviation in energy
discrepancy between this tight-binding model and the two interatomic
potentials are $0.40\pm0.04$~eV and $0.57\pm0.07$~eV, with the
Stillinger-Weber yielding the lower value.  Accordingly, we use the
Stillinger-Weber potential to explore the phase space of our system.

\begin{figure}
\begin{center}
\PSbox{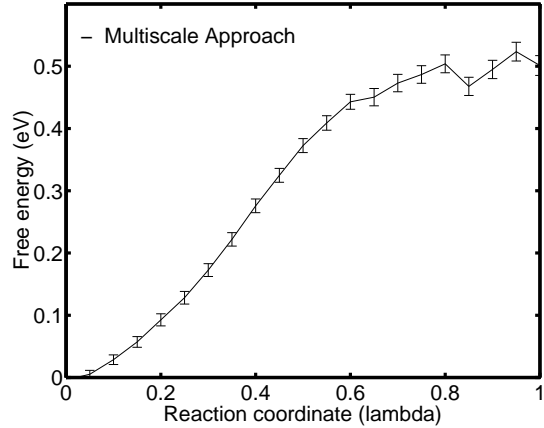 hoffset=-15  voffset=-80 vscale=40 hscale=40}{3.3in}{2.0in}
\end{center}
\caption{Multiscale calculation of quantum mechanical free energy of
reconstruction}
\label{fig:tbsw}
\end{figure}

Figure~\ref{fig:tbsw} presents the quantum mechanical tight-binding
free energy profile of reconstruction calculated with the multiscale
approach.  Our results are based on the energy and forces of only
twelve {\em hundred} samples drawn from the Stillinger-Weber
potential.  With the traditional Monte Carlo approach, a run of such
length would produce at most a few independent points in phase space,
while our calculation represents an extended exploration of the
canonical distribution.  Our tight-binding result for the free energy
of reconstruction is $0.53\pm0.015$~eV per broken bond in the
unreconstructed dislocation.

This value falls within limits placed by available experimental 
information and lends support to mechanisms for
dislocation mobility suggested by atomistic studies \cite{APD}.  
In these atomistic studies, the energy required to
break the reconstructed bonds in the dislocation core contributes
significantly to the activation energy for dislocation
mobility.  The Stillinger-Weber value for the reconstruction energy,
$0.81$~eV, however, is too high in that it leads
to a prediction \cite{Bulatov_private} for the dislocation mobility
which is multiple orders of magnitude lower than
observed experimentally \cite{George_Champier}.  Our value for the {\em free} 
energy includes explicitly both thermal fluctuations and bonding
effects.  The fact that it is significantly lower than the
Stillinger-Weber value tends to increase the predicted dislocation
mobility and thus lends support to the mechanisms proposed in
 \cite{APD}.  Our lower free energy leads also to a significantly
higher prediction for the equilibrium density of dangling bonds in the
dislocation core.  The total signal strength in electron paramagnetic
resonance experiments \cite{Alexander}, whose precise origin is
difficult to interpret and may involve other defects, places an
absolute upper limit of a few percent on the density of dangling bonds
associated with the dislocations.  Our free energy value corresponds
to a dangling bond density of $\sim 0.1\%$, well below the
experimental bound. 

{\em Conclusions} --- There is tremendous benefit in separating
thermal studies of a system with highly detailed models into
two parts: an exploration of phase space using a simpler
coarse-grained Hamiltonian and use of a more detailed Hamiltonian to
study the behavior of the system at a limited number of well-chosen
points in phase space.
We have seen that the approach works well for the determination of the
free energies of defects from quantum-based calculations.  As system
size increases, the determination of such local defects is relatively
stable.  Global changes, such as phase transitions, in which the
observable of interest is correlated with the entire system may
require a different approach.  A thermal, quantum mechanical
description of the $30^o$ partial dislocation core gives a free energy
of reconstruction which is significantly lower than previous
Stillinger-Weber values and, thus, leads to a more consistent view of
dislocation mobility in silicon.
Finally, the number of samples required to apply the multiscale
approach to systems of approximately one hundred atoms is sufficiently
low as to make its application to {\em ab initio} calculations
attractive.

\begin{center}{\bf Acknowledgments}\end{center}
Financial support: LLNL (B332671), MRSEC Program of the NSF (DMR
94-00334), Alfred P. Sloan Foundation (BR-3456), Research Council of
Norway (115898/410).  Computational support: MIT Xolas prototype SUN cluster.


\begin{thebibliography}{0}
\bibitem{TerC}J. Tersoff, {\em Phys. Rev. Lett.} {\bf 61}, 2879 (1988).

\bibitem{SW}F.Stillinger and T.Weber, {\em Phys. Rev.} {\bf B31},
5262 (1985). 

\bibitem{Ter}J.Tersoff, {\em Phys. Rev.} {\bf B38}, 9902 (1988).

\bibitem{Daw12}M.S. Daw and M.I. Baskes, {\em Phys. Rev. Lett.} {\bf
50}, 1285 (1983). 

\bibitem{Harrison}W.A. Harrison, {\em Surface Science}, {\bf 299-300}
290 (1994).

\bibitem{Sawada}S. Sawada, {\em Vacuum}, {\bf 41}, 612 (1990).

\bibitem{Kohyama}M. Kohyama, {\em J. Phys.: Cond. Matt.}, {\bf 3},
2193 (1991).


\bibitem{Krajci}M. Kraj\v{c}\'{\i} and J. Hafner, {\em
Phys. Rev. Lett.} {\bf 74}, 5100 (1995).

\bibitem{RMP} M.C. Payne {\em et al.},
{\em Rev. Mod. Phys.} {\bf 64}, 1045 (1992).

\bibitem{Milman}V. Milman, {\em et al.} {\em Phys. Rev. Lett.} {\bf
70}, 2928 (1993).

\bibitem{MD}L. Verlet {\em Phys. Rev.}, {\bf 159}, 98 (1967).

\bibitem{Me}Metropolis {\em et al.}, {\em J. Chem. Phys}, {\bf 21},
1087 (1953). 

\bibitem{umbrella}J. P. Valleau and D. N. Card, {\em J. Comp. Phys.}{\bf
23}, 187 (1977).

\bibitem{FI}J. G. Kirkwood, {\em J. Chem. Phys.} {\bf 3}, 300 (1935).

\bibitem{George_Champier}A. George and G. Champier, {\em
Phys. Stat. Sol.} {\bf A53}, 529 (1979).
 
\bibitem{Alexander}H. Alexander in {\em Dislocations in Solids},
F.R.N. Nabarro, Ed. (Elsevier, Amsterdam, 1986), Vol. 7, Ch. 35,
p. 126.

\bibitem{Becke}A.D. Becke, {\em J. Chem. Phys.}, {\bf 98}, 5648
(1993). 

\bibitem{Bulatov_private}V.V. Bulatov, {\em private communication}.

\bibitem{APD}V. Bulatov,  S. Yip,  A.S. Argon, {\em Phil. Mag.}, 453 (1995).
\end{thebibliography}
\end{document}